\documentclass[12pt,epsf,fleqn]{article}
\usepackage{epsfig}
\usepackage{color}
\begin{document}
\title{Higgs bosons of a supersymmetric $U(1)'$ model at the ILC}
\author{S. W. Ham$^{(1)}$ \footnote{Email address: {\sf swham@knu.ac.kr}},
E. J. Yoo$^{(2)}$, S. K. Oh$^{(1,2)}$, and D. Son$^{(1)}$
\\
\\
{\it $^{(1)}$ Center for High Energy Physics, Kyungpook National University}
\\
{\it Daegu 702-701, Korea}
\\
{\it $^{(2)}$ Department of Physics, Konkuk University, Seoul 143-701, Korea}
\\
}
\date{ }
\maketitle
\begin{abstract}
We study the scalar Higgs sector of the next-to-minimal supersymmetric standard model with an extra $U(1)$,
which has two Higgs doublets and a Higgs singlet, in the light leptophobic $Z'$ scenario
where the extra neutral gauge boson $Z'$ does not couple to charged leptons.
In this model, we find that the sum of the squared coupling coefficients of the three neutral scalar Higgs bosons to $ZZ$,
normalized by the corresponding SM coupling coefficient is noticeably smaller than unity,
due to the effect of the extra $U(1)$, for a reasonable parameter space of the model,
whereas it is unity in the next-to-minimal supersymmetric standard model.
Thus, these two models may be distinguished if the coupling coefficients of neutral scalar Higgs bosons to $ZZ$
are measured at the future International Linear Collider by producing them via the Higgs-strahlung, $ZZ$ fusion,
and $WW$ fusion processes.
\end{abstract}
%
%
\vfil
\eject

\section{Introduction}

Despite all its success against precision tests so far, the standard model (SM) is widely considered
as the low-energy effective approximation of a fundamental theory.
Two of the major directions to the fundamental theory beyond the SM may be via supersymmetry or grand unification,
as numerous theoretical attempts to formulate it have been accumulated in these directions.

The search for supersymmetric extensions of the SM is motivated by the No-Go theorem, the gauge hierarchy problem,
and the possibility of incorporating gravity by local SUSY [1].
Supersymmetric models extended from the SM require at least two Higgs doublets, in order to give masses
to the up-quark sector and the down-quark sector separately.
While the minimal version has just two Higgs doublets [2], nonminimal versions may have
one or more Higgs singlets besides the two doublets.
The next-to-minimal supersymmetric standard model (NMSSM) is the simplest one of those nonminimal versions,
which has one additional Higgs singlet [3].
The NMSSM has five independent neutral Higgs fields, which may be separated into three scalar
and two pseudoscalar Higgs bosons if CP is conserved.

Grand unification is required in order to unify interactions in a consistent way,
to reduce free parameters of the SM, and to explain naturally the fermion mass problem [4,5].
In $SO(10)$, for example, the chiral fermion fields of the SM are unified
into three generations of an irreducible representation.
A number of Lie groups have been investigated within the context of grand unification.
The mainstream of the grand unifications are such that their gauge groups are broken down,
possibly several times, at higher scales, to yield the SM gauge symmetry as a subgroup.
At the electroweak scale, the true character of the grand unification may be investigated
in retrospect if some structure of the unification group persist to remain.
In other words, for instance, if the grand unification, after its breaking,
gives rise to $SU(2)\times U(1) \times U(1)'$ at the electroweak scale,
the extra $U(1)'$ might lead the the original unification group.

In recent days, supersymmetric grand unifications have also been studied
in the literature for the new physics beyond the SM.
These studies have become more interesting by the recent numerical observation that,
if one assumes the existence of superpartners of the SM particles,
the unification of three renormalization-group gauge coupling coefficients works particularly well.
The NMSSM may also be incorporated with grand unification such that
at the electroweak scale the gauge symmetry of the NMSSM may accommodate an additional $U(1)$
to become $SU(2)\times U(1) \times U(1)'$.
The string-inspired $E_6$ model might be a natural motivation
to introduce the extra $U(1)$ symmetry to the NMSSM.

The NMSSM with the extra $U(1)$ (UNMSSM) may differentiate itself from the NMSSM by exhibiting distinct phenomena [6,7].
The existence of $Z'$ is a typical signal that distinguishes the two models.
Since the mass of $Z'$ is generally expected to be larger than about 600 GeV,
establishing experimental evidences for $Z'$ would take some time.
Meanwhile, $G_{ZZS_i}$ ($i =1,2,3$), the coupling coefficients of the neutral scalar Higgs bosons to $ZZ$,
may provide a clue to distinguish the two models.
In the NMSSM, there is a sum rule for these coupling coefficients [8].
We find that the corresponding sum rule is altered in the UNMSSM significantly.
It is the purpose of this article to study the phenomenology of the NMSSM with this extra $U(1)$,
which we may call the UNMSSM afterwards, and to discuss how to distinguish between them.

At the future $e^+ e^-$ International Linear Collider (ILC) with $\sqrt{s} = 800$ GeV,
we anticipate that the neutral scalar Higgs bosons would most dominantly be produced
through the $WW$ fusion process,
but also through the Higgs-strahlung process and the $ZZ$ fusion process.
We would like to show that we may tell, once the mass of at least one of
the three neutral scalar Higgs bosons and its coupling coefficient to a $ZZ$ pair are determined,
whether the neutral scalar Higgs boson belong to the UNMSSM or
to the NMSSM .

\section{UNMSSM}

We assume that the extra $U(1)$ emerges from $E_6$.
In other words, a low-energy superstring-inspired $E_6$ model is assumed to give rise
to $SU(2)\times U(1) \times U(1)'$ of the UNMSSM.
As Green and Schwarz has shown some years ago, string theories in ten dimensions can
be anomaly free at the string-theory level if the gauge group is either $E_8 \times E_8^{'}$ or $SO(32)$ [5].
Between the two gauge groups, $E_8 \times E_8^{'}$ can be reduced to $E_6$
as an effective grand unification at low energy by compactifying additional six dimensions,
and it can accommodate chiral fermions.

The $E_6$ may be decomposed into
\begin{equation}
    E_6 \supset SO(10)\times U(1)_{\psi} \supset SU(5) \times U(1)_{\chi} \times U(1)_{\psi} \ ,
\end{equation}
where $SU(5)$ is further broken down to the SM gauge group, $SU(3)_C \times SU(2)_L \times U(1)_Y$.
At the electroweak scale, the desired extra $U(1)$ symmetry may be given
as an orthogonal linear combination of $U(1)_{\chi}$ and $U(1)_{\psi}$ as
\begin{equation}
    U(1)' = \cos \theta_E U(1)_{\chi} + \sin \theta_E U(1)_{\psi} \ .
\end{equation}
The electroweak gauge group we consider is thus $G = SU(3)_C \times SU(2)_L \times U(1)_Y \times U(1)'$.
Sometimes, particular values of the mixing angle $\theta_E$ designate specific model names as follows:
the $\chi$-model for $\theta_E = 0$, the $\psi$-model for $\theta_E = \pi/2$,
the $\eta$-model for $\theta_E = \tan^{-1} (-\sqrt{5/3})$,
and the $\nu$-model for $\theta_E = \tan^{-1} \sqrt{15}$.
We take the $\eta$-model.

The structure of the neutral Higgs sector of the UNMSSM is identical to that of the NMSSM:
There are two Higgs doublets $H_1 = (H_1^0, H_1^-)$ and $H_2 = (H_2^+, H_2^0)$, and a neutral Higgs singlet $S$.
There are therefore ten real degrees of freedom.
After spontaneous symmetry breaking, the three neutral Higgs fields develop vacuum expectation values
as $\langle H_1^0 \rangle = v_1$, $\langle H_2^0 \rangle = v_2$, and $\langle S \rangle = s $,
where we assume that CP is conserved.
For simplicity, we take only the third generation of quarks into account.
Then, the superpotential for the Yukawa interactions in the UNMSSM for quarks
and the exotic quarks may be expressed as
\begin{equation}
W \approx h_t Q^T \epsilon H_2 t_R^c + h_b Q^T \epsilon H_1 b_R^c + h_k S k_L {\bar k}_R + \lambda H_1^T \epsilon H_2 S  \ ,
\end{equation}
where $\epsilon$ is an antisymmetric $2 \times 2$  matrix with $\epsilon_{12} = 1$,
and $h_t$, $h_b$ and $h_k$ are respectively the dimensionless Yukawa coupling coefficients of top, bottom, and exotic quarks,
$t_R^c$ and $b_R^c$ are the right-handed top and bottom quark superfields, respectively,
$Q$ is the left-handed $SU(2)$ doublet quark superfield of the third generation,
and the right and left handed singlet exotic quark superfields are denoted respectively as $k_R$ and $k_L$.
Notice that there is a cubic interaction among the three Higgs multiplets
with a dimensionless coupling coefficient $\lambda$.

The low-energy particle content of an $E_6$ grand unification model can be contained
in the fundamental 27 representation of $E_6$,
where the SM matter particles occupy 15 components, the two Higgs doublets occupy 4 components,
and the Higgs singlet occupies 2 components.
The exotic quarks are introduced in order to complete the fundamental 27 representation of $E_6$.
Thus, the remaining 6 components of the fundamental 27 representaion are occupied by the exotic quarks.

Let us denote the effective hypercharges of $H_1$, $H_2$, and $S$,
respectively, as  ${\tilde Q}_1$, ${\tilde Q}_2$, and ${\tilde Q}_3$.
They are defined as
\[
    {\tilde Q}_i = Q_i^{\eta} + \delta Q_i^Y
\]
($i =1, 2, 3$), where $Q_i^Y$ are the $U(1)_Y$ hypercharges of the corresponding Higgs fields,
$\delta $ is the kinetic mixing parameter, and $Q^{\eta}_i$ are given by
\[
    Q^{\eta}_i = \cos \theta_E Q^{\chi}_i + \sin \theta_E Q^{\psi}_i \ ,
\]
where $Q^{\eta}_i$, $Q^{\chi}_i$, and $Q^{\psi}_i$ are the hypercharges of $U(1)'$, $U(1)_{\chi}$,
and $U(1)_{\psi}$, respectively.
These hypercharges satisfy $ {\tilde Q}_1 + {\tilde Q}_2 + {\tilde Q}_3  = 0 $
by virtue of the $U(1)'$ gauge invariance of the superpotential under $U(1)'$.

The kinetic mixing parameter $\delta$ is given as  $\delta = g_{11}/g'_1$,
where $g'_1$ is the  $U(1)'$ gauge coupling constant and $g_{11}$ is
a new gauge coupling coefficient arising from the mixing between $g_1$ and $g'_1$.
Without mixing between $g_1$ and $g'_1$, $g_{11}$ as well as $\delta$ would be zero and ${\tilde Q}_i$ would reduce to $Q'_i$.
At the electroweak scale, the analysis of renormalization group equation yields that $g'_1(m_Z) \sim 0.46$ [9].
It is observed that, since the coupling of $Z'$ to a charged lepton pair is proportional
to the effective hypercharge of the charged lepton, $Z'$ would not couple to charged leptons
if the effective hypercharges of charged leptons could be rendered zero [10].
This leptophobic condition for $Z'$ is satisfied if $\delta = 1/3$.
In our analysis, we assume that $Z'$ is leptophobic.

The extra neutral gauge boson $Z'$ is essentially the core characteristic of the UNMSSM.
The mixing between $Z$ and $Z'$ is described in terms of a mixing angle $\phi$, which is given by
\begin{equation}
\phi = {1 \over 2} \tan^{-1} \left ({ 2 \Delta^2 \over  m_{Z'}^2-m_Z^2} \right ) \ ,
\end{equation}
where $m_Z = (g_1^2+g_2^2) v^2 /2$ is the $Z$ mass,
$m_{Z'}^2  = 2 {g'_1}^2 v^2 ({\tilde Q}_1^2 \cos^2 \beta + {\tilde Q}_2^2 \sin^2 \beta ) + 2 {g'_1}^2 {s^2 \tilde Q}_3^2$ is the $Z'$ mass,
and $\Delta^2 = \sqrt{g_1^2 + g_2^2} g'_1 v^2 ({\tilde Q}_1 \cos^2 \beta - {\tilde Q}_2 \sin^2 \beta)$.
Note that $\phi$ is not a free parameter but it depends on $g'_1$, ${\tilde Q}_i$ ($i=1,2,3$), $\tan \beta$ and $s$.

Both the $Z'$ mass and the mixing angle $\phi$ in the UNMSSM are strongly constrained by experiments.
According to the latest Review of Particle Physics, the lower bound on the $Z'$ mass
in the $\eta$ model is about 745 GeV from $p{\bar p}$ direct search [11] and 619 GeV by electroweak fit [12].
It is remarked that, if the $Z'$ is leptophobic,
that the experimental sensitivities are much weaker,
and searches for a $Z'$ via hadronic decays at CDF are unable to rule out a $Z'$
with quark couplings identical to those of $Z$ in any mass region [13].
For instance, as Fig. 3 in Ref. [14] suggests, the $Z'$ mass may be as small as 100 GeV.
From a phenomenological point of view, we are interested in a light $Z'$.
A light $Z'$ appears when the vacuum expectation value of the Higgs singlet, $s$, is about 700 GeV.
A general comment on $s$ is that larger $s$ leads to heavier $Z'$ in the present model.

It is constrained by precision measurements that $\phi$ in the UNMSSM should be smaller than 2-3 $\times 10^{-3}$.
This constraint on $\phi$ may also be loosened if the $Z'$ is leptophobic such that the mixing angle might be as large as 0.06 [14].
Anyway, a large $s$ satisfies the experimental constraint on $\phi$ in the leptophobic case.

\section{Numerical analysis}

Let us take a reasonable parameter space for numerical analysis.
We take $1 < \tan \beta \leq 30$, $0 < \lambda \le 0.85$, and $700 \leq s \leq 2000$ GeV,
where $\tan\beta = v_2/v_1$, and $v = \sqrt{v^2_1 +v^2_2} = 175$ GeV.
The $U(1)'$ gauge coupling constant is taken as $g'_1= 0.46$, motivated by the gauge coupling unification.
The SUSY breaking mass, $m_{\rm SUSY}$, arising from the soft SUSY breaking terms,
is assumed to be within the range from 100 GeV to 1000 GeV.

The three neutral scalar Higgs bosons, denoted as $S_i$ ($i=1,2,3$), are obtained
as the eigenstates of the $3\times 3$ mass matrix for three neutral real Higgs fields.
The upper bound on $m_{S_1}$, the mass of the lightest neutral scalar Higgs boson $S_1$,
at the one-loop level in the UNMSSM, including radiative corrections due to the quarks
and squarks of the third generation, is given by [7]
\begin{eqnarray}
m_{S_1}^2 & \le & \lambda^2 v^2 \sin^2 2 \beta + m_Z^2 \cos^2 2 \beta \cr
& &\mbox{}+ 2 {g'_1}^2 v^2 ({\tilde Q}_1 \cos^2 \beta + {\tilde Q}_2 \sin^2 \beta)^2 + \Delta m_{S_1}^2 \ ,
\end{eqnarray}
where $\lambda$ is the dimensionless coupling coefficient for the cubic Higgs interaction,
and $\Delta m_{S_1}^2$ is the radiative correction, which comes from the one-loop corrections
due to the quarks and squarks of the third generation.
It is a function of $s$.
Among the relevant parameters, the upper bound on $m_{S_1}$ depends significantly
on the values of $\lambda$ and ${\tilde Q}_i$.

Numerical calculation shows that the upper bounds on $m_{S_1}$ becomes maximum
at $\tan \beta \sim 1.5$ and then decreases as $\tan\beta$ increases.
This is mainly because the effect of the Higgs singlet and the $D$-term contribution due to the extra $U(1)$
at the tree-level Higgs potential interfere each other.
This behavior is different from the MSSM, where the upper bound on $m_{S_1}$ increases as $\tan \beta$ increases.
We find that $m_{S_1} \le 162$ GeV in the parameter ranges of our choice
in the leptophobic scenario of the UNMSSM, at the one-loop level.

Now, let us study $G_{ZZS_i}$ ($i = 1,2,3$), the coupling coefficients of the neutral scalar Higgs bosons to $ZZ$.
In the SM, the corresponding quantity is given as
\[
    G_{ZZH} = g_2 m_Z/\cos \theta_W  \ ,
\]
where $\theta_W$ is the weak mixing angle, and $H$ is the neutral scalar Higgs boson of the SM.
We may normalize $G_{ZZS_i}$ by $G_{ZZH}$ such that ${\bar G}_{ZZS_i} = G_{ZZS_i}/G_{ZZH}$.
Explicitly, the normalized coupling coefficients may be written as
\begin{equation}
{\bar G}_{ZZS_i} (\phi) = \cos \beta O_{1i} C_1^2 + \sin \beta O_{2i} C_2^2 + {s O_{3i} \over 4 G_{ZZH}} C_3^2 \ ,
\end{equation}
where $O_{ij}$ ($i,j =1,2,3$) are the elements of an orthogonal transformation matrix
that diagonalizes the $3 \times 3$ mass matrix for the three neutral scalar Higgs bosons,
and $C_i$ ($i = 1,2,3$) are dimensionless parameters defined by
\begin{eqnarray}
C_1 & = & \cos \theta_W^2 + \sin \theta_W^2 \cos \phi
- {g'_1 {\tilde Q}_1 \over g_2} \cos \theta_W \sin \theta_W \sin \phi \ , \cr
C_2 & = & \cos \theta_W^2 + \sin \theta_W^2 \cos \phi
+ {g'_1 {\tilde Q}_2 \over g_2} \cos \theta_W \sin \theta_W \sin \phi \ , \\
C_3 & = & g'_1 {\tilde Q}_3 \sin \theta_W \sin \phi \ . \nonumber
\end{eqnarray}
Notice that in the above expressions the $Z$-$Z'$ mixing angle $\phi$ is present in $C_i$,
which is responsible for the interference effect of the extra gauge boson.

If $\phi = 0$, we would have $C_1= C_2 =1$ and $C_3=0$, and ${\bar G}_{ZZS_i} (0) = \cos \beta O_{1i} + \sin \beta O_{2i}$,
which are exactly the normalized coupling coefficients of $S_i$ to $ZZ$ in the NMSSM.
The mixing angle $\phi$ is indeed zero if $Z'$ decouples from $Z$, or if the extra $U(1)$ is absent in the UNMSSM.
In the latter case, the UNMSSM reduces trivially to the NMSSM.

Now, we would like to remark that ${\bar G}_{ZZS_i} (0)$,
the three normalized coupling coefficients in the NMSSM, satisfy a simple sum rule:
\begin{equation}
    \sum_{i=1}^3 {\bar G}_{ZZS_i}^2(0) = 1 \ .
\end{equation}
On the other hand, ${\bar G}_{ZZS_i} (\phi)$, the normalized coupling coefficients in the UNMSSM,
would not satisfy the sum rule, since $\phi$ would not vanish in general.
Because of the non-zero $\phi$, we would have
\begin{equation}
\sum_{i=1}^3 {\bar G}_{ZZS_i}^2 (\phi) = \cos^2 \beta C_1^4 + \sin^2 \beta C_2^4
+ \left ({s C_3^2 \over 4 G_{ZZH} } \right )^2 \ne 1 \ .
\end{equation}
and thus non-trivial mixing between $Z$ and $Z'$.
Therefore, by examining the value of $\sum_{i=1}^3 {\bar G}_{ZZS_i}^2 (\phi)$,
one may in principle detect the effect of the existence of $Z'$,
and thus distinguish the UNMSSM from the NMSSM.

\begin{figure}[t!]
\begin{center}
\includegraphics[scale=0.6]{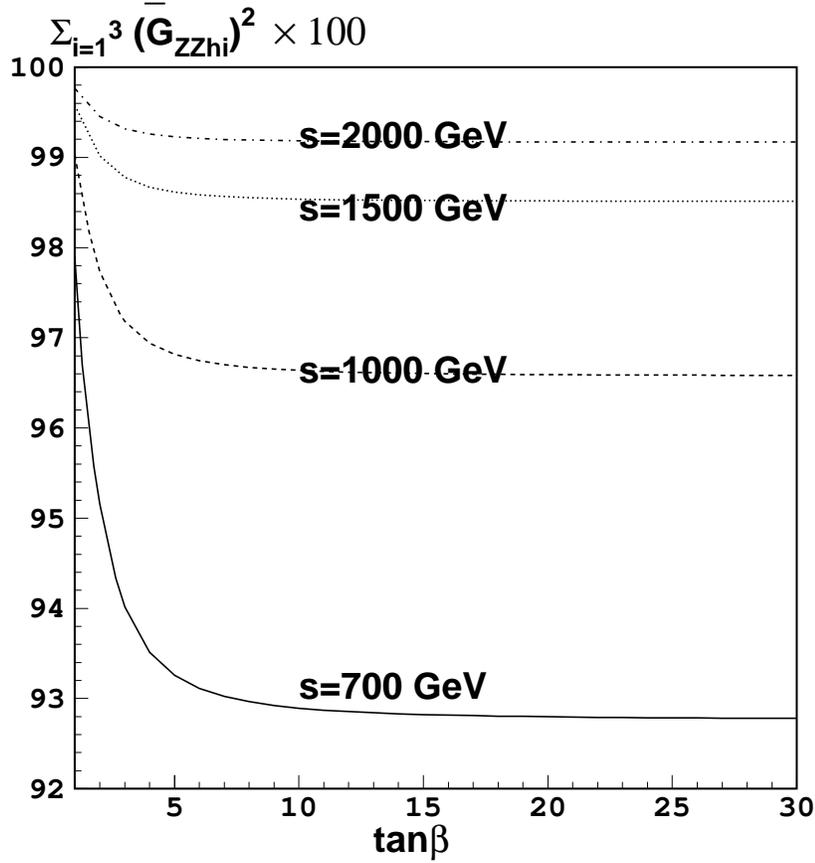}
\caption{$\sum_{i=1}^3 {\bar G}_{ZZS_i}^2 (\phi)$ against $\tan \beta$ in the UNMSSM,
for four different values for $s$: 700 GeV (solid curve),
1000 GeV (dashed curve), 1500 GeV (dotted curve), and 2000 GeV (dotted-dashed curve).}
\end{center}
\end{figure}
\begin{figure}[t!]
\begin{center}
\includegraphics[scale=0.6]{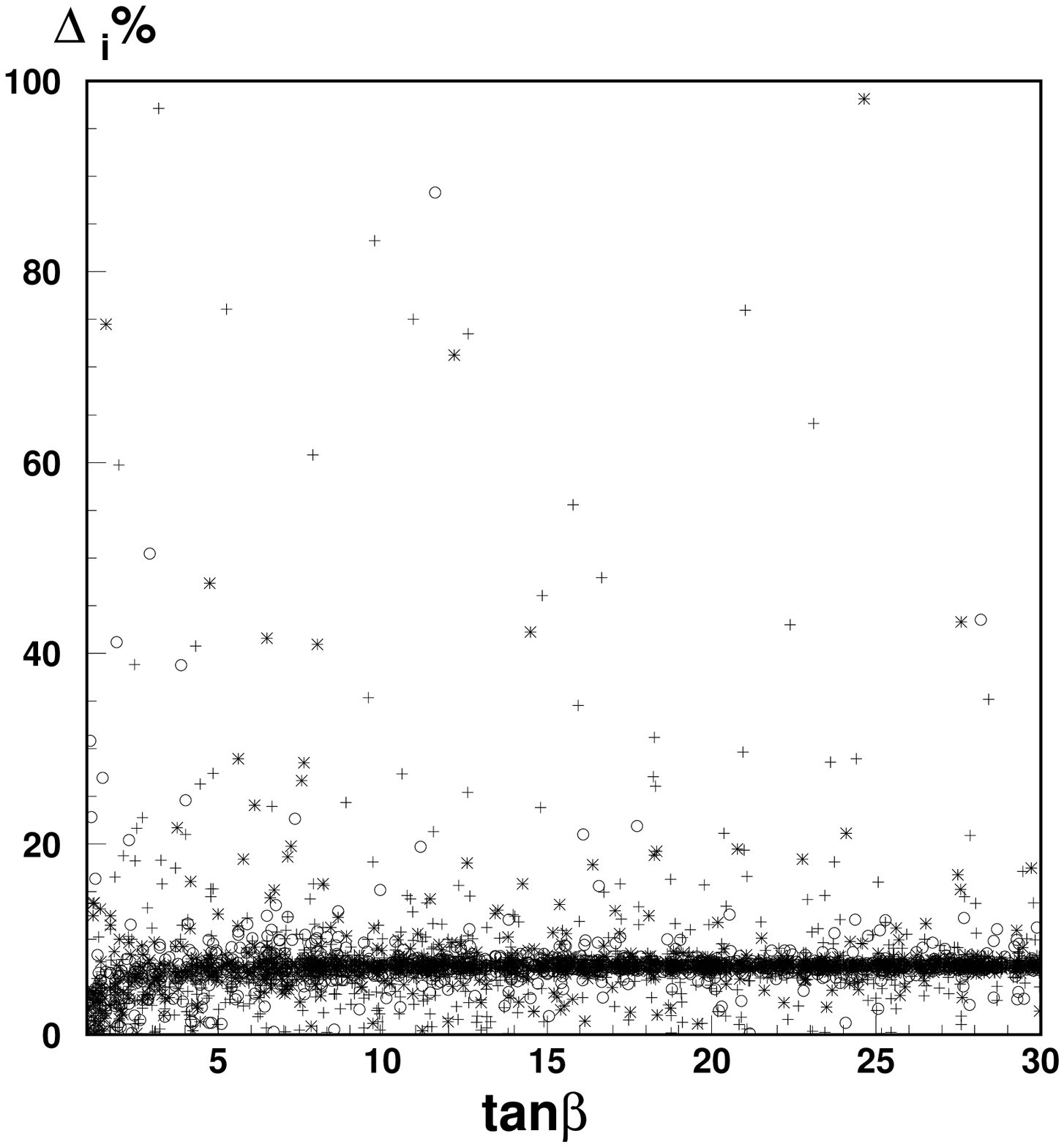}
\caption{$\Delta_i$ against $\tan \beta$ for $s=700$ GeV in the UNMSSM,
where parameter values are randomly varied within the allowed ranges.
The points marked by star, circle, and cross are respectively $\Delta_1$, $\Delta_2$, and $\Delta_3$.}
\end{center}
\end{figure}

For numerical analysis, we need to know the values of $O_{ij}$.
They are generated randomly by varying the relevant parameters within the allowed ranges,
with the constraint that the matrix $O$ is an orthogonal matrix.
Then, we evaluate $\sum_{i=1}^3 {\bar G}_{ZZS_i}^2 (\phi)$.
In Fig. 1, we plot the result for $\sum_{i=1}^3 {\bar G}_{ZZS_i}^2 (\phi)$ against $\tan \beta$,
for four different values for $s$: 700 GeV (solid curve),
1000 GeV (dashed curve), 1500 GeV (dotted curve), and 2000 GeV (dotted-dashed curve).
It is quite clear that $\sum_{i=1}^3 {\bar G}_{ZZS_i}^2 (\phi) \ne 1$ in the UNMSSM,
for the given ranges of parameters.
The difference is more vivid for larger $\tan \beta$ and for smaller $s$.

Please note that, for $s=700$ GeV, we obtain $|\phi| <$ 0.03-0.037 and $644 \le m_{Z'} \le 650$ GeV,
depending on $\tan \beta$.
Our choice of the parameter values yields the $ZZ'$ mixing and the $Z'$ mass
within their experimental constraints in the leptophobic case [14].

One may quantify the difference between the UNMSSM and the NMSSM by introducing
\[
\Delta_i = { |{\bar G}_{ZZS_i}^2(\phi) - {\bar G}_{ZZS_i}^2(0)| \over {\bar G}_{ZZS_i}^2(0)} \ .
\]
In Fig. 2, we plot $\Delta_i$ against $\tan \beta$ for $s=700$ GeV,
where parameter values are randomly varied within the allowed ranges.
The points marked by star, circle, and cross are respectively $\Delta_1$, $\Delta_2$, and $\Delta_3$.
For most of the parameter space, $\Delta_i$ are about 7 \%, which is quite recognizable.
The plot of Fig. 2, as well as the result of Fig. 1,
suggests that the neutral scalar Higgs sector of the UNMSSM may clearly be distinguished from that of the NMSSM.

Let us now study productions of these neutral scalar Higgs bosons in $e^+e^-$ collisions at the ILC,
where the center of mass energy of the $e^+e^-$ system is sufficiently high in order to produce them.
The important channels to produce them are the Higgs-strahlung process $e^+ e^- \rightarrow Z S_i$,
the $WW$ fusion process $e^+e^- \rightarrow {\bar \nu}_e \nu_e S_i $, and the $ZZ$ fusion process $e^+e^- \rightarrow e^+ e^- S_i$.
We denote the production cross sections via each of these processes in the UNMSSM
as $\sigma^H_i (\phi)$, $\sigma^W_i (\phi)$, and $\sigma^Z_i (\phi)$, respectively.

These production cross sections are related to the cross sections for the production of the SM Higgs boson via the corresponding process.
For the Higgs-strahlung process, $\sigma^H_i (\phi)$ in the UNMSSM is related to the production cross section for the SM Higgs boson
\[
\sigma_i^H (\phi) = (G_{ZZS_i}(\phi)/ G_{ZZH})^2 \sigma^H_{\rm SM} =  {\bar G}_{ZZS_i}^2(\phi) \sigma^H_{\rm SM}  \ .
\]
Likewise, for the $ZZ$ fusion process,
\[
\sigma_i^Z (\phi) = {\bar G}_{ZZS_i}^2(\phi) \sigma^Z_{\rm SM}
\]
Similar relationship may be established between the production cross sections in the NMSSM and those in the SM:
\begin{eqnarray}
&&\sigma_i^H (0) = {\bar G}_{ZZS_i}^2(0) \sigma^H_{\rm SM} \\
&&\sigma_i^Z (0) = {\bar G}_{ZZS_i}^2(0) \sigma^Z_{\rm SM} \nonumber
\end{eqnarray}

Note that, however, for the $WW$ fusion process,
the effect of the extra neutral gauge boson is absent and we may set $\phi = 0$.
Thus, ${\bar G}_{WWS_i}^W (\phi) = {\bar G}_{WWS_i}^W (0)$.
Further, it holds that $G_{WWS_i} (0) /G_{WWH} = {\bar G}_{ZZS_i}(0)$, where $G_{WWH}$ is
the coupling coefficient of the SM Higgs boson to $WW$.
Consequently, we have, for the $WW$ fusion process,
\[
    \sigma_i^W (0) = {\bar G}_{ZZS_i}^2 (0) \sigma^W_{\rm SM}
\]
both in the UNMSSM and in the NMSSM.

{\color{red}
\setcounter{table}{0}
\def\tablename{}{}%
\renewcommand\thetable{TABLE 1}
\begin{table}[t]
\caption{The masses, the normalized coupling coefficients to $ZZ$,
and the production cross sections of the three neutral scalar Higgs bosons via Higgs-strahlung, $ZZ$ fusion
and $WW$ fusion process in the UNMSSM, for $\tan \beta=10$, $m_Q=500$ GeV, $\lambda=0.35$, $A_t=500$,
and $m_p=200$ GeV at the ILC with $\sqrt{s}=800$ GeV.
Those quantities with $\phi =0$ are the NMSSM values, except for $\sigma^W_i(0)$,
which are also the UNMSSM values: $\sigma^W_i(0) = \sigma^W_i(\phi)$.}
\begin{center}
\begin{tabular}{c|c|c|c} \hline\hline
i & 1 & 2 & 3  \\
\hline\hline
$m_{S_i}$ GeV & 140 & 201 & 295 \\
\hline
${\bar G}_{ZZS_i}^2 (\phi)$ & 0.9018 & 0.0162 & 0.0108    \\
\hline
${\bar G}_{ZZS_i}^2 (0)$ & 0.9699 & 0.0185 & 0.0114    \\
\hline
$\sigma_i^W (0)$ fb & 62.68 & 0.620 & 0.103  \\
\hline
$\sigma_i^H (\phi)$ fb & 18.02 & 0.297 & 0.156   \\
\hline
$\sigma_i^H (0)$ fb & 19.38 & 0.336 & 0.166   \\
\hline
$\sigma_i^Z (\phi)$ fb & 6.362 & 0.060 & 0.010   \\
\hline
$\sigma_i^Z (0)$ fb & 6.843 & 0.068 & 0.011   \\
\hline\hline
\end{tabular}
\end{center}
\end{table}
}

At the ILC energy, the $WW$ fusion process would be the most dominant production channel
in $e^+e^-$ collisions for the lightest neutral scalar Higgs boson.
Once the existence of a neutral scalar Higgs boson is established at the ILC,
its nature will be examined to determine whether it is the SM Higgs boson or
the lightest neutral scalar Higgs bosons in the NMSSM or in the UNMSSM.
If it turns out to be the latter one, the experiments would determine
its mass $m_{S_1}$ and the coupling coefficient ${\bar G}_{ZZS_1}^2(0)$.

These experimentally measured values would then constrain the free parameters of each model.
In terms of ${\bar G}_{ZZS_1}^2(0)$, one may obtain $\sigma_1^Z (0)$ and $\sigma_1^H (0)$,
which are the production cross sections of $S_1$ in the NMSSM.
On the other hand, in terms of ${\bar G}_{ZZS_1}^2(\phi)$,
one may also obtain $\sigma_1^Z (\phi)$ and $\sigma_1^H (\phi)$,
which are the production cross sections of $S_1$ in the UNMSSM.
Such a situation is summarized in Table 1.

Notice that all of the three neutral Higgs bosons are not so heavy for the parameter values we take
and thus they are kinematically within the reach of the ILC.
It is evident from Table 1 that the production cross sections
in the UNMSSM are predicted to be reasonably smaller than those in the NMSSM,
due to the interference effect by the extra neutral gauge boson.
Also note that $\sum_{i=1}^3 {\bar G}_{ZZS_i}^2(0) = 1$,
whereas $\sum_{i=1}^3 {\bar G}_{ZZS_i}^2(\phi) < 1$.
These numbers may then be compared with the corresponding data at the ILC.
Practically, it would be difficult to measure all these quantities during the first stages of operations at the ILC.
Nevertheless, by comparing ${\bar G}_{ZZS_1}^2(0)$ with ${\bar G}_{ZZS_1}^2(\phi)$ in Table 1,
finding a distinction between the UNMSSM and the NMSSM would be possible.
In particular, the sum rule for the coupling coefficients may easily be examined.
Any discrepancy of the sum rule from unity would distinguish the UNMSSM from the NMSSM,
by confirming in an indirect way the existence of the extra neutral gauge boson.

\section {Conclusions}
The extra neutral gauge boson in the UNMSSM is expected to manifest its existence
in the Higgs sector of the model.
There are three neutral scalar Higgs bosons in the UNMSSM.
The mass of the lightest scalar Higgs boson is constrained from above by, among others,
the effective $U(1)'$ charges of the Higgs doublets and the Higgs singlet.
The upper bound on it is about 162 GeV, at the one-loop level,
which suggests that it can be discovered at the ILC.
For certain parameter values of the UNMSSM, such as those in Table 1,
it is kinematically possible that the ILC may produce all of the three neutral Higgs bosons if they exist.

The existence of the extra $U(1)'$ gauge group induces the mixing effect between $Z$ and $Z'$
in the production processes of these neutral scalar Higgs bosons.
It reduces the sum of the squared coupling coefficients of
the three neutral scalar Higgs bosons to $ZZ$, in particular the coupling coefficient of
the lightest neutral scalar Higgs boson,
such that the distinction between the UNMSSM and the NMSSM is practicable.
We suggest that if a neutral scalar Higgs boson is produced via the Higgs-strahlung,
$ZZ$ fusion, or $WW$ fusion processes are measured at the ILC,
one may be able to examine whether it is identified as the lightest one in the UNMSSM or in the NMSSM.

\vskip 0.3 in
\noindent
{\large {\bf ACKNOWLEDGMENTS}}
\vskip 0.2 in
This work is KOSEF through CHEP, Kyungpook National University.
The authors would like to acknowledge the support from KISTI (Korea Institute of Science
and Technology Information) under "The Strategic Supercomputing Support Program"
with Dr. Kihyeon Cho as the technical supporter.
The use of the computing system of the Supercomputing Center is also greatly appreciated.


\end{document}